\documentstyle[12pt,epsfig,sprocla]{article}
\begin{document}
~
\vskip -2.0 cm
{\hskip 10.0 cm \bf DFUB 98/22}
\begin{center}
\vskip 1.0 cm
{\bf ATMOSPHERIC NEUTRINOS AND NEUTRINO OSCILLATIONS}
\vskip .5 cm
{G.GIACOMELLI and M.SPURIO}
\vskip .3 cm
{\it Dipartimento di Fisica dell'Universit\`a di Bologna and INFN, Sezione 
di Bologna, 40127 Bologna, Italy}

{\it E-mail:giacomelli@bo.infn.it , spurio@bo.infn.it}
\end{center}
\vskip .3 cm
\noindent {\small Lecture at the Fifth School on Particle Astrophysics, 
Trieste 29 june-10 july 1998}
\vskip 1.5 cm
\begin{abstract}
After some generalities on neutrino oscillations and on neutrinos, the
recent experimental  results presented by Soudan 2, MACRO and SuperKamiokande
at the Neutrino'98 conference are summarized and discussed.
\end{abstract}

\section{Introduction}
\begin{figure}
 \vspace{-1cm}
 \begin{center}
  \mbox{ \epsfysize=7cm
         \epsffile{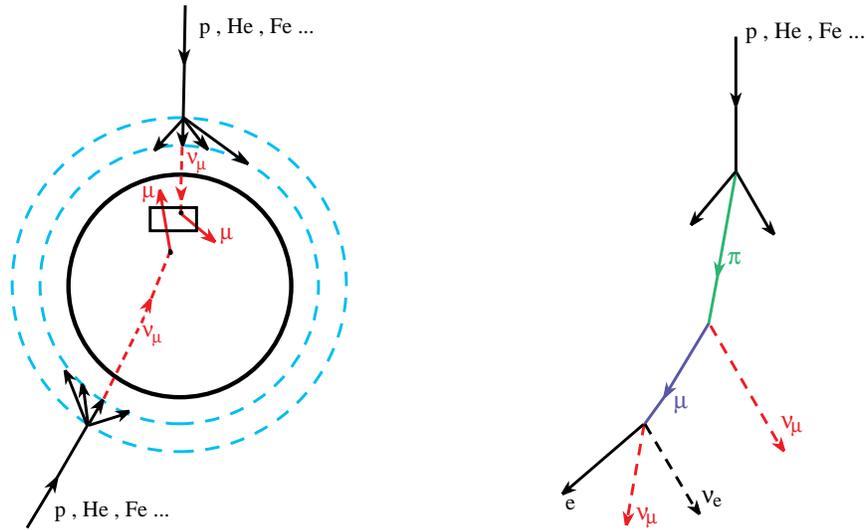} }
 \end{center}
\caption{\label{fig:prof}\small 
(a) Illustration
of the production, travel and interactions of atmospheric
muon neutrinos; (b) interaction of a primary
cosmic ray, production of pions and their decays
leading to the atmospheric $\nu_e,\nu_\mu$.}
\end{figure}

Atmospheric neutrinos are produced in cosmic ray interactions in the 
upper atmosphere: a high energy primary cosmic ray, either proton or nucleus, 
interacts producing  a large number of hadrons, in
particular pions and kaons. These can decay giving rise to muons and muon
neutrinos; also the muons decay yielding muon and electron neutrinos.
In this simplified picture, the ratio of the
numbers of muon to electron
neutrinos is 2, 
$(N_{\nu_\mu} + N_{\overline\nu_\mu})/(N_{\nu_e} +  N_{\overline\nu_e}) 
\simeq 2 $,
and $(N_{\nu}/N_{\overline\nu}) \simeq 1$, see Fig. \ref{fig:prof}.
One may consider that these neutrinos are produced in a spherical
surface at about few tens of km above ground, 
and that they proceed at high velocity
towards the center of the earth.

The neutrino flux has been computed by a number of authors.
At low energies $E_\nu \sim \ 1\ GeV$, the absolute numbers of predicted
neutrinos differ by about $30\%$\cite{nulow};
at higher energies, $E_\nu > 10\ GeV$, the predictions are more
reliable, with a systematic  uncertainty of about $15\%$, that is
about $1/2$ of that at low energies\cite{nuhigh}. However the predicted
relative rates of $\nu_\mu$ and $\nu_e$ and the shape of the zenith 
distribution have a
considerably lower systematic error. Other sources of systematic
uncertainties are due to the knowledge of the 
neutrino-nucleon cross sections and to the propagation of muons and
electrons in different materials.

Since the 1980's several large underground detectors, mainly designed to 
search for proton decays, have studied atmospheric neutrinos. 
These detectors are located below 1-2 km of rock and they may detect 
neutrinos coming from all directions. Via charged current ($CC$) 
interaction the 
$\nu_\mu$ gives rise to $\mu^-$ and thus to a track,
the $\nu_e$ to $e^-$ and thus to an electromagnetic shower.
The produced hadronic system is not observed in most cases. 

The detectors can be roughly classified as tracking calorimeters and as
water Cherenkov detectors. The water Cherenkov detectors 
Kamiokande\cite{kamioka} and IMB\cite{imb} reported 
anomalies in the ratio of muon to electron
neutrinos, while the  tracking calorimeters NUSEX\cite{nusex} and
Frejus\cite{frejus} did not find any deviation.
Also the Baksan\cite{baksan} scintillator detector did not 
see any deviation. 

Later Soudan 2\cite{soudan} 
and MACRO\cite{macro} reported
deviations; in 1997 the neutrino anomaly was also reported by 
SuperKamiokande\cite{skam}.

At the Session of June 5, 1998 at the Neutrino'98 Conference in Takayama,
Japan, new, higher statistics data have been presented by the 
Soudan 2\cite{sou98}, MACRO\cite{mac98} and SuperKamiokande\cite{sk98} 
collaborations. The experiments confirmed the
neutrino anomaly and opened a wide
discussion on the subject.

The main purpose of this lecture is to review and discuss the
presentations at the Neutrino'98 Conference. The SuperKamiokande results have
been presented at this School by M. Koshiba\cite{koshiba}.
Some results have been updated at the 1998 HEP Conference 
in Vancouver, Canada\cite{conrad}.

\section{ Neutrino oscillations}

If neutrinos have non-zero masses one has to consider 
$\nu_e, \nu_\mu ,\nu_\tau$ and their combinations. The
$\nu_e, \nu_\mu ,\nu_\tau$ are the appropriate particles to consider in
weak decays, for example $\pi^+ \rightarrow \mu^+ + \nu_\mu$,
and in charged current ($CC$)
interactions which lead to their detection, {e.g.}
$\nu_\mu + n \rightarrow \mu^- + p$; in technical
terms one says that they are {\it weak flavour eigenstates}.
Instead in the propagation in vacuum the appropriate particles
are the  {\it mass eigenstates} $\nu_1,\ \nu_2,\ \nu_3$.
The weak flavour eigenstates $\nu_l$ are linear combinations of the mass 
eigenstates $\nu_m$:
\begin{equation}
\nu_l = \sum_{m=1}^3 U_{lm}\ \nu_m
\end{equation}
In the simplest case of only two neutrinos $(\nu_\mu,\nu_\tau)$ which
oscillate with two mass eigenstates $(\nu_2,\nu_3)$ one may write
\begin{equation}
\begin{array}{ll}
          \nu_\mu =~\nu_2\ cos\ \theta_{\mu\tau} + \nu_3\ sin\ \theta_{\mu\tau} \\
          \nu_\tau=-\nu_2\ sin\ \theta_{\mu\tau} + \nu_3\ cos\ \theta_{\mu\tau} 
\end{array} 
\end{equation}

\noindent where $\theta_{\mu\tau}$ is the mixing angle.

If the mixing angles are small one would have
$\nu_e \sim \nu_1$,
$\nu_\mu \sim \nu_2$,
$\nu_\tau \sim \nu_3$; in this case one may speak of the mass of $\nu_e$
which is about equal to that of $\nu_1$, etc.
In the limit of zero masses the neutrinos become equal, and one
does not need to introduce the $\nu_1, \nu_2,\nu_3$.
If the mixing angles are large, the situation is different
and one has to consider well separated the eigenstates
 of flavour from those of mass.
In particular it would be inappropriate to speak of
mass for the weak eigenstates $\nu_e, \nu_\mu ,\nu_\tau$. 

In the case of only two types of neutrinos, $\nu_\mu$ and
$\nu_\tau$, one may easily compute the following expression for the survival
probability of a $\nu_\mu$ beam:
\begin{equation}
P(\nu_\mu \rightarrow \nu_\mu) = 1- sin^2 2\theta_{\mu\tau}  
\ sin^2 ( { {E_2-E_1}\over {2}} t) =
1- sin^2 2\theta_{\mu\tau}\ sin^2 ( { {1.27 \Delta m^2 \cdot L}\over {E_\nu}})
\end{equation}

The probability for the initial $\nu_\mu$ to have oscillated into a
$\nu_\tau$ is:
\begin{equation}
P(\nu_\mu \rightarrow \nu_\tau) = 1 - P(\nu_\mu \rightarrow \nu_\mu) =
 sin^2 2\theta_{\mu\tau}\  sin^2 ( { {1.27 \Delta m^2 \cdot L}\over {E_\nu}})
\end{equation}
The mixing angle $\theta_{\mu\tau}$ and the mass difference $\Delta m^2 
= \Delta m^2_{\nu_2\nu_3}$ may be determined from the variation of
$ P(\nu_\mu \rightarrow \nu_\mu) $ as a function of 
the zenith angle $\Theta$, or the path $L$, or the energy $E_\nu$ or
$L/E_\nu$.

\section{ Early experiments}

The ring imaging water Cherenkov detectors Kamiokande
and IMB measured $\nu_\mu$ 
and $\nu_e$ $CC$ interactions and found that the ratio of muons
to electrons was smaller than expected\cite{stone}. In the  Kamiokande
experiment, neutrino interactions were classified using the shape
of the Cherenkov rings on the phototubes on the wall of the 
cylindrical water container, and
through the recognition of muon decays. The results were
expressed in terms of the ratio $R=R_{obs}/R_{MC}$ between 
$R_{obs} = ({{\nu_\mu} \over {\nu_e}})_{obs}$ of measured $CC$ interaction
events and $R_{MC} = ({{\nu_\mu} \over {\nu_e}})_{MC}$ from Monte Carlo 
simulations. The single ratio  $(\nu_\mu)_{obs}/(\nu_\mu)_{MC}$,
may be affected by large theoretical and systematic uncertainties; in the 
double ratio most systematic uncertainties cancel.
The measured double ratios from Kamiokande\cite{kamioka} and IMB\cite{imb} are 
$R_{Kamioka} = 0.60\pm 0.06_{stat+sys}$ and $R_{IMB} = 0.60\pm 0.06_{stat}
\pm 0.07_{sys}$, respectively.
The NUSEX\cite{nusex} and Frejus\cite{frejus} tracking calorimeter detectors
reported for contained and semicontained events $R_{obs}\sim R_{MC}$
within errors. The  measured double ratios are: $R_{Nusex} = 1.0\pm 0.3$,  
$R_{Frejus} = 0.99\pm 0.13_{stat} \pm 0.08_{sys}$. 

The Baksan scintillation telescope detected a sizable number of
upthroughgoing muons arising from $\nu_\mu$ interactions in the
rock below the detector\cite{baksan}. The
average $\nu_\mu$ energy for these events is considerably larger 
$(50-100\ GeV)$ than for the contained events measured by the other detectors.
They did not find deviations from the predictions in the total number of events,
but they find an anomalous angular distribution.

Later, the Soudan 2 tracking and shower  calorimeter confirmed an anomaly
in the $\nu_\mu/\nu_e$ ratio for contained events\cite{soudan}.

The MACRO\cite{macro} collaboration reported in 1995 a first measurement
of upthroughgoing muons coming from $\nu_\mu$ of 
$\overline E_\nu\sim \ 100\ GeV $
in which there was a deficit in the total number of observed upgoing
muons, in particular around the vertical, and also reported an 
anomalous zenith angle distribution. The deficit was confirmed 
at the 1996 and 1997 conferences\cite{macro}.

At the 1997 conferences, the SuperKamiokande Collaboration confirmed
the Kamiokande and IMB anomalies
in the $\mu/e$ ratio for contained events\cite{skam}.

\section{ Contained events and Soudan 2}

The Soudan 2 
data were presented at 
Neutrino'98 by E. Peterson\cite{sou98}; improved data were presented at HEP'98
by H. Gallagher\cite{sou98}.
The Soudan 2 experiment uses a modular fine grained tracking and showering
calorimeter of $963\ t$. It is located $2100\ m.w.e.$ underground in the
Soudan Gold mine in Minnesota, USA.
Its overall dimensions are $8m \times 16m \times 5m$. The
{\it Central Detector} 
is made of 224 calorimetric modules of $1m \times 1m \times 2.5m$. The
bulk of the mass consists of 1.6-mm-thick corrugated  steel sheets which are
interleaved with drift tubes.
The detector is surrounded by an anticoincidence. This 
{\it Active Shield} detector covers the walls of the Soudan 2 cavern enclosing
the Central Detector as hermetically as possible.

The neutrino contained events
were selected by a combination of a two-stage software
filter and a two-stage physicist scan.
The software filter rejects non-contained events by requiring
that {\it(i)} no part of the event is within 20 cm of the
detector surface and {\it(ii)} no track is located
or oriented in such a way that it could enter undetected in the 
calorimeter from a crack between modules.
The last stage is the scan of real and simulated events. The
MC simulated events were mixed with the real ones. 
All events were classified into one of
three topologies: tracks (as due to $\nu_\mu\ CC$ interactions);
showers ($\nu_e\ CC$ interactions) and multiprongs (interactions
of all neutrino flavors and $NC$). The multiprong events are not
considered at present.
Finally, events without any hits in the Active Shield detector 
are called {\it Gold Events}, while events with two or more
hits in the shield are called {\it Rock Events}.
 
\begin{table}
\begin{center}
\begin{tabular}
{cccccc}\hline
 & Gold Data & Rock data & Background & Events & MC \\ \hline
Track     &  95 & 278 & $18.5\pm 6.1$ & $76.9\pm10.8$  & 137.4\\
Showers   & 151 & 473 & $33.3\pm12.8$ & $116.3\pm12.8$ & 133.8\\ \hline
\end{tabular}
\end {center}
\caption {\small Event summary for the Soudan 2 data.
Track events are due to $\nu_\mu$ while showers to $\nu_e$. The
MC predictions were obtained using the $\nu$ Bartol flux.} 
\label{tab:soudan}
\end{table}

The data from a $3.89\ kt\cdot yr$ exposure are summarized in
Table \ref{tab:soudan}, together with the Monte Carlo predictions
using the Bartol neutrino flux\cite{Agrawal96}.
The Soudan 2 double ratio 
is $R=(N_\mu/ N_e)_{data}/ (N_\mu/ N_e)_{MC} 
= 0.64 \pm 0.11 ^{+0.06}_{-0.05}$
which is consistent with muon neutrino oscillations.

\begin{figure}
 \begin{center}
  \mbox{ \epsfysize=7.5cm
         \epsffile{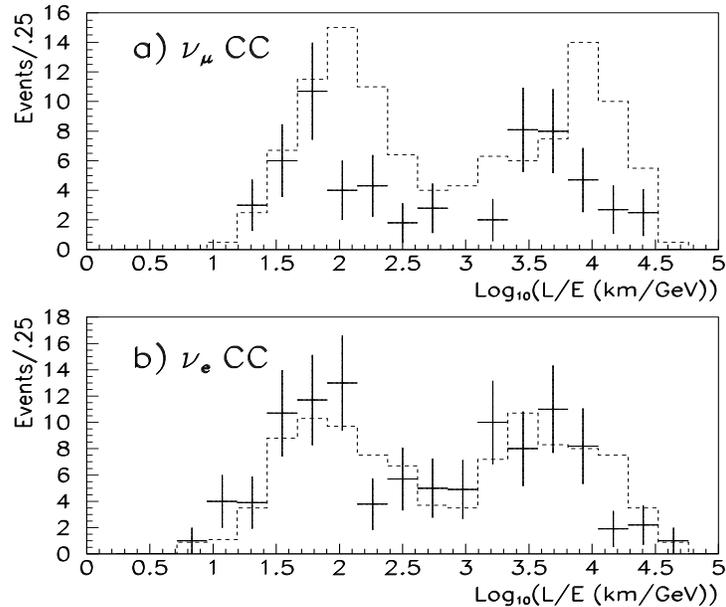} }
 \end{center}
\caption {\label{fig:sres}\small 
Soudan 2 (preliminary) results on the number of 
observed events as function of 
$L/E$: (a) for $\nu_\mu$ and (b) $\nu_e$. Only statistical errors are shown; 
the dashed lines are the MC predictions using the Bartol flux.}
\end{figure}

For a smaller, high resolution sample, they are able to estimate the
$L/E$ for each event (Fig. \ref{fig:sres}), 
together with the MC predictions. 
Within limited statistics (after corrections, $60.8\ \nu_\mu\ CC$ events
and $106.4 \ \nu_e\ CC$ events), 
the preliminary
data are consistent with an anomaly in the muon data 
and not in the electron data. For this data set $R= 0.52\pm 0.09_{stat}$.
They add that $\Delta m^2 < 10^{-3}\ eV^2$
appears unlikely.

\section{Upward-going muons and MACRO}

\begin{figure}
 \vspace{-1cm}
 \begin{center}
  \mbox{ \epsfysize=8cm
         \epsffile{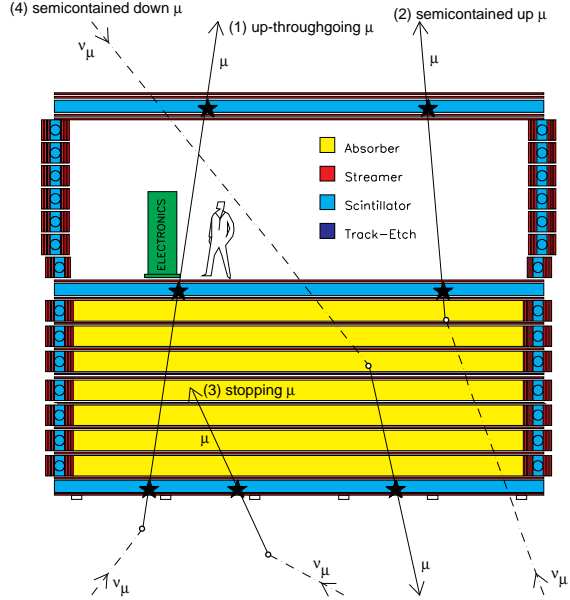} }
 \end{center}
\caption {\label{fig:topo}\small Sketch of
different event topologies induced
by $\nu_\mu$
interactions in or around MACRO. The stars
represent scintillator hits. The time-of-flight of the particle 
can be measured for {\it Up Semicontained} and {\it Up throughgoing} events.}
\end{figure}

The MACRO data were presented at Neutrino'98 by F.Ronga\cite{mac98}.
The MACRO detector is located in Hall B of the Gran Sasso
Laboratory, with a minimum rock overburden of 3150 hg/cm$^2$.
It is a large rectangular box,
76.6~m~$\times$~12~m~$\times$~9.3~m, divided
longitudinally in six similar supermodules and vertically in a lower
part (4.8 m high) and an upper part (4.5 m high). The
detection elements are planes of streamer tubes for tracking
and liquid scintillation counters for fast timing. The lower half of
the detector is filled
with trays of crushed rock absorbers alternating with streamer tube
planes; the upper part is open and contains the
electronics. There are 10
horizontal planes in the bottom half of the detector, and 4 planes on the
top with wire and 27$^\circ$ stereo strip readouts.  Six vertical planes
of streamer tubes cover each side of the detector.
The scintillator system consists of three layers 
of horizontal counters, and of one vertical layer
along the sides of the detector.  The time (position) resolution for
muons in a scintillation counter is about 500~ps (11~cm).
Fig. \ref{fig:topo} shows 
a sketch of the three different topologies of neutrino
events analyzed until now: up throughgoing muons, semicontained
upgoing muons and up stopping muons+semicontained downgoing muons.

The {\it  up throughgoing muons} come from $\nu_\mu$ 
interactions in the rock below the detector,
with $\overline E_\nu \sim \ 100\ GeV$. The muons $(E_\mu > 1\ GeV)$
cross the whole detector. The time information 
provided by scintillation counters
allows the determination of the direction by the time-of-flight (T.o.F.)
method. 
The data presented correspond to $\sim 3.5$ live years.

\begin{table}
\begin{center}
\begin{tabular}
{cccc}\hline
 & Events & \multicolumn{2}{c}{Predictions (Bartol neutrino flux)}  \\ \cline{3-4}
 & detected & No Oscillations & With oscillations \\ \hline
Up Through       & 451 & $612\pm 104_{th} \pm 37_{sys}$ & $431\pm 73_{th} \pm 26_{sys}$ \\
Internal Up      &  85 & $144\pm  36_{th} \pm 14_{sys}$ & $83\pm 21_{th} \pm 8_{sys}$ \\
In Down+Up Stop  & 120 & $159\pm  40_{th} \pm 16_{sys}$ & $123\pm 31_{th} \pm 12_{sys}$ \\
\hline
\end{tabular}
\end {center}
\caption {\small Event summary for the MACRO analysis. The predictions with
oscillations are for maximum mixing and 
$\Delta  m^2 = 0.0025\ eV^2$.} 
\label{tab:macro}
\end{table}

The {\it semicontained upgoing muons} come from
$\nu_\mu$ interactions inside the lower apparatus. Since two
scintillation counters are intercepted, the T.o.F.
is applied to identify the upward going muons. The average 
parent neutrino energy for these events is $\sim 4\ GeV$. 
If the atmospheric neutrino anomalies are the results of
$\nu_\mu$ oscillations with maximum mixing
and $\Delta m^2$ between $10^{-3}$ and $10^{-2}\ eV^2$ one expects a 
reduction of about a factor of two in the flux of these events, 
without any distortion in the shape of the angular distribution.

The {\it up stopping muons}
are due to external $\nu_\mu$ interactions yielding upgoing muon tracks
stopping in the detector; the {\it semicontained downgoing muons} are due to 
$\nu_\mu$ induced downgoing tracks with vertex in the lower MACRO.
The events are found by means of topological criteria; the lack
of time information prevents to distinguish the two sub samples.
An almost equal number of  up stopping  and  semicontained downgoing 
events is expected, and the average neutrino energy for these events is
around $4\ GeV$.
In case of oscillations with the quoted parameters,
a similar reduction in the flux of the up stopping events 
as the semicontained upgoing muons is expected. No reduction is 
instead expected for the semicontained downgoing events (from
neutrinos having  path lengths of $\sim 20\ km$).

\begin{figure}
 \vspace{-2.5cm}
 \begin{center}
  \mbox{ \epsfysize=8cm
         \hspace{-1cm}
         \epsffile{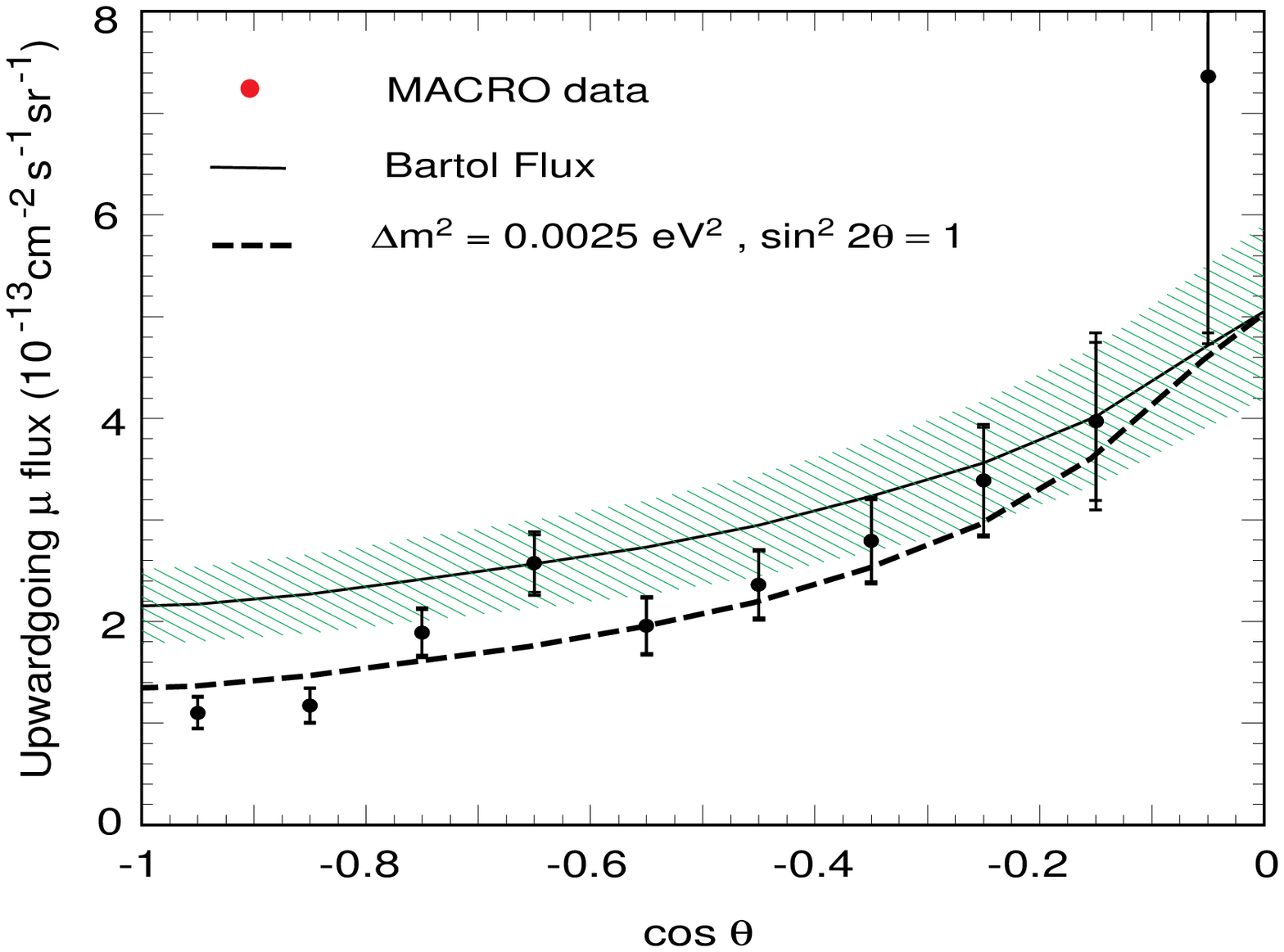}    \hspace{-0.7cm}
         \epsfysize=6.5cm
         \vspace{3.5cm}
         \epsffile{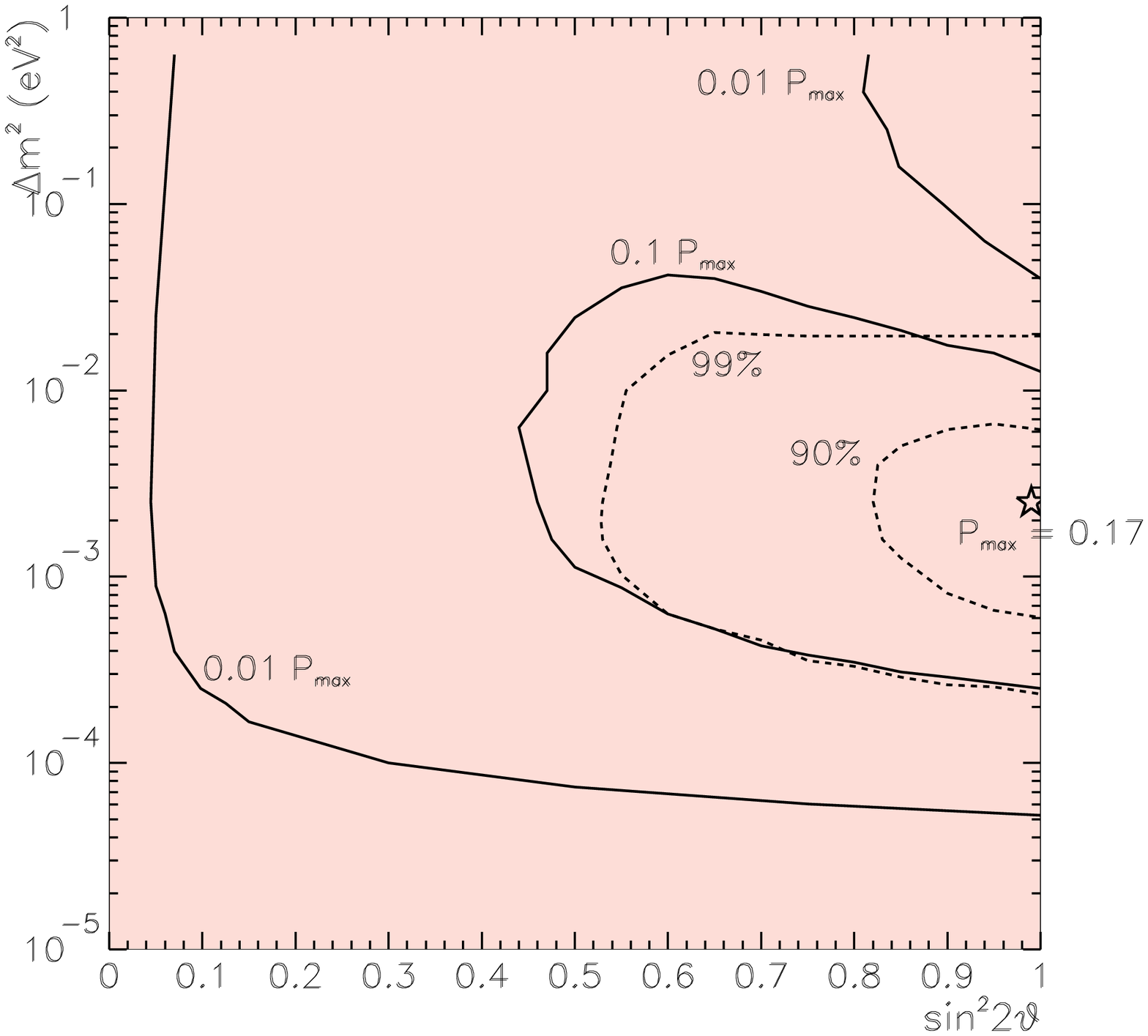} }
 \end{center}
\caption{\label{fig:cosze}\small  MACRO data. (a) Flux of the
up throughgoing muons with $E_{\mu} > 1$ GeV vs. zenith angle $\Theta$.
The solid line is the expectation for no
oscillations and the shaded region is a 17\% scale 
uncertainty. The dashed line is the prediction 
for an oscillated flux with  maximum mixing and
$\Delta m^{2} = 0.0025$ eV$^{2}$.
(b) Probability contours 
for $\nu_\mu \rightarrow \nu_\tau$ oscillations. 
The dashed lines are 90\% and 99\% CL contours calculated according
to$^{21}$.
The best probability is 17\%; iso-probability contours are
shown for 10\% and 1\% of this value (i.e. 1.7\% and 0.17\%).}
\end{figure}

The data were compared with Monte Carlo simulations. 
In the upgoing muon simulation 
the neutrino flux computed by the Bartol group\cite{Agrawal96} is used.
The cross sections for the neutrino interactions have been calculated
using the  Morfin and Tung\cite{Morfin91}  parton distribution set S1.
The propagation of muons to the detector has been done using the
energy loss calculation by Lohmann {\em et al.}\cite{Lohmann85}
in standard rock. The total systematic uncertainty from
neutrino flux, cross section and muon propagation on the
expected flux of muons is $\sim$17\%. 
Fig. \ref{fig:cosze}a   shows the zenith angle distribution of the measured
flux of up throughgoing muons with energy greater than 1 GeV;
 the Monte Carlo expectation for no oscillations is shown as a solid line, and
for a $\nu_\mu \rightarrow \nu_\tau$ oscillated flux with
$\sin^2 2\theta =1$ and  $\Delta m^2= 0.0025\ eV^2$ is 
shown by the dashed line.
The systematic uncertainty on the  up throughgoing muons flux is 
mainly a scale error that doesn't change the shape of the angular distribution.
The  ratio of the observed
number of events to the expectation without oscillations is
$0.74\ \pm 0.036_{stat}\ \pm 0.046_{sys}\ \pm 0.13_{theor}$.

The shape of the angular distribution of Fig. \ref{fig:cosze}a has 
been tested with the hypothesis of no oscillation, giving a $\chi^2$ of
26.1 for 8 degrees of freedom.
Assuming $\nu_\mu \rightarrow \nu_\tau$ oscillations, the best $\chi^2$
in the physical region of the oscillations parameters is $15.8$ for
$\Delta  m^2=0.0025\ eV^2$ and $\sin^2 2\theta=1$.

 To test oscillation hypotheses,
the independent probability for
obtaining the number of events observed and the angular distribution
for various parameter values have been calculated.
The value of $\Delta  m^2$ suggested from the shape of the
angular distribution is similar to the value
needed to obtain the observed reduction in the number of events in the
hypothesis of maximum mixing. Fig.  \ref{fig:cosze}b  shows probability
contours for oscillation parameters using the combination of
probability for the number of events and the 
$\chi^2$ of the angular distribution.
The maximum probability is 17\%. The probability for no
oscillations is 0.1\%.

\begin{figure}
 \vspace{-2cm}
 \begin{center}
  \mbox{ \epsfysize=7cm
         \hspace{-1cm}
         \epsffile{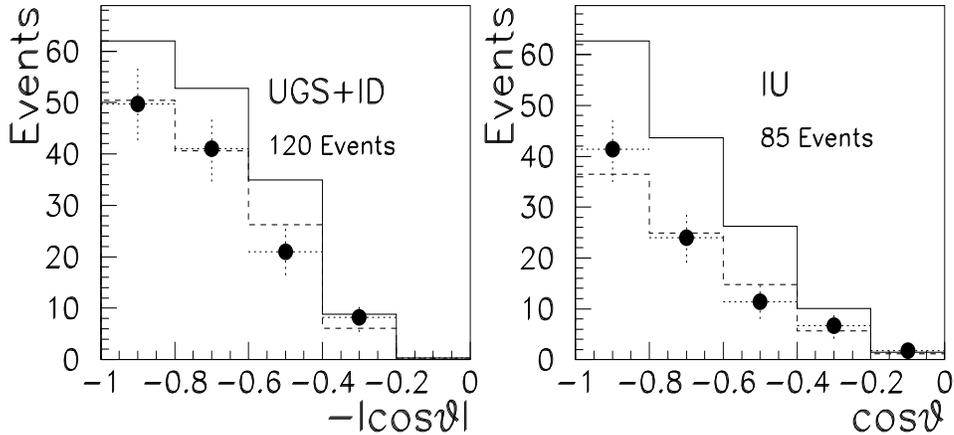} }
 \end{center}
\caption {\label{fig:low_cosze}\small 
MACRO data. Measured and expected number of low energy events versus 
zenith angle; left: up stopping plus down semicontained;
right: up semicontained.
The solid lines are the predictions
without oscillations; the dashed lines are the predictions assuming neutrino 
oscillations with the parameters suggested by the Up throughgoing
sample.}
\end{figure}

The MC simulation for the low energy data uses the Bartol neutrino
flux and the neutrino low energy cross sections of ref.\cite{lipari94}.
The number of events and the angular distributions are compared with the
predictions in Table \ref{tab:macro} and Fig. \ref{fig:low_cosze}. The low
energy data show a uniform deficit on the measured number of events
over the whole angular distribution with respect to the predictions;
there is good agreement with the predictions based on neutrino
oscillations using the parameters obtained from the up throughgoing muon
sample.

\section{Results from SuperKamiokande and Kamiokande}

\begin{figure}
 \begin{center}
  \mbox{ \epsfysize=6cm
         \epsffile{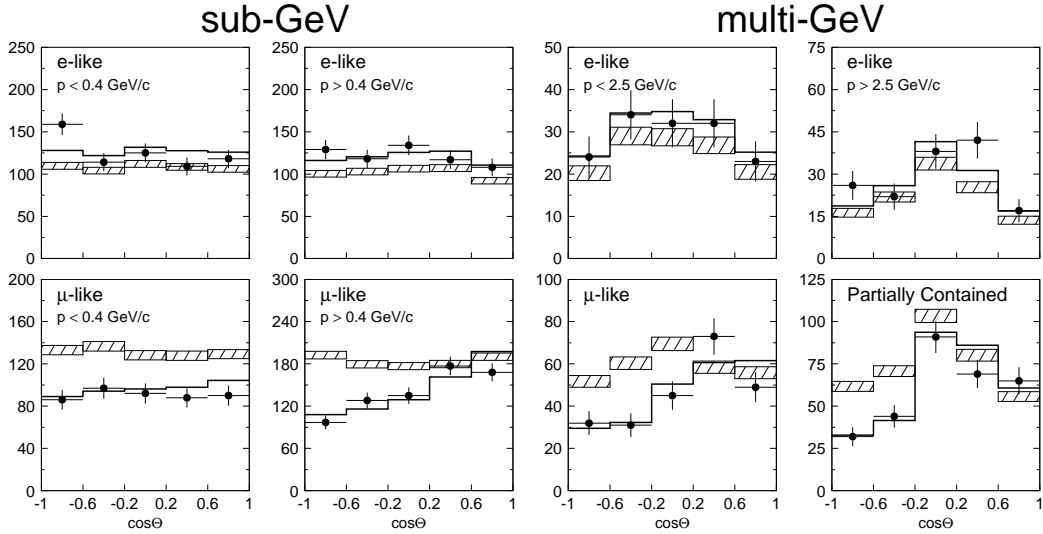} }
 \end{center}
\caption{\label{fig:sk3}\small 
SuperKamiokande data. 
Zenith angle distributions of $\mu$-like and $e$-like events
for sub-GeV and multi-GeV data sets. Upward- going particles have
cos$\Theta <0$. The hatched regions are the Monte Carlo expectation 
for no oscillations
normalized to the detector live-time. The solid lines are 
the best-fit expectations for 
$\nu_\mu \rightarrow \nu_\tau$ oscillations with the overall flux
normalization fitted as  free parameter.}
\end{figure}

SuperKamiokande is a large cylindrical water Cherenkov detector 
of  $39\ m$ diameter and  $41\ m$ height containing $50000\ m^3$ of 
water. The inner detector of $22500\ m^3$ is seen by 11146, 50-cm-diameter 
inner-facing phototubes. The $2 m$ thick outer layer of water acts 
as an anticoincidence and is seen by 1885 smaller outward-facing
photomultipliers. The ultra pure water
has a light attenuation of almost $100\ m$. The detector 
is located in the Kamioka mine, Japan, under $2700\ m.w.e.$. For more
details see the lecture by M. Koshiba\cite{koshiba}.

Because of the water index of reflection of 1.33, a relativistic charged
particle (as the muon generated by a neutrino $CC$ interaction) generates
a $42^o$ forward cone of light. Instead electrons, being
much lighter than muons, suffer
electromagnetic showering and multiple scattering; therefore the ring
of light is not so well defined as for the muons. The detector
can distinguish single muons from single electrons
with about 98\% efficiency. Multiple ring events have been
only partially used. The light
intensity on the phototubes is a measure of the particle energy,
and its direction is determined by the spatial
and temporal pattern of phototube hits.

The data presented at Neutrino'98 correspond to $33\ kt\cdot yr$ 
of data. This exposure collected 4353 fully contained
(FC) events and 301 partially contained (PC) events.
The FC events were separated into "sub-GeV" (with $E_{vis}<1.3\ GeV)$
and "multi-GeV" ($E_{vis}>1.33\ GeV)$ samples; $E_{vis}$ is defined to be the
energy of an electron that would produce the observed amount
of Cherenkov light. $E_{vis}=1.33\ GeV$ corresponds to $p_\mu \sim 1.4 \ GeV/c$.

In a full-detector Monte Carlo simulation, 88\% (96\%) of sub-GeV
$e$-like ($\mu$-like) events were $\nu_e$ ($\nu_\mu$) charged-current
interactions and 84\% (99\%) of the multi-GeV $e$-like ($\mu$-like) 
events were $\nu_e$ ($\nu_\mu$) $CC$ interactions.
PC events were estimated to be 98\% $\nu_\mu$ charged-current
interactions; hence, all the PC events were classified as $\mu$-like,
and no single-ring requirement was made.

\begin{figure}
 \vspace{-3cm}
 \begin{center}
  \mbox{ \epsfysize=8cm
         \hspace{0.5cm}
         \epsffile{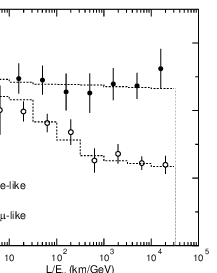} \hspace{0.5cm}
         \epsfysize=8cm
         \epsffile{fig9.eps} }\vspace{0.5cm}
 \end{center}
\caption{\small
SuperKamiokande data: 
(a) The ratio of the number of FC events to
Monte Carlo events versus reconstructed
$L/E_{\nu}$.
The points are the ratios of observed data to MC expectation
in the absence of oscillations. The lower dashed line
is the expected shape for $\nu_\mu \rightarrow \nu_\tau$
for $\Delta m^2 = 2.2 \times 10^{-3}$ and  $sin^22\theta =1$.
(b) The 68\%, 90\% and 99\% C.L. contours
for $sin^22\theta$ and  $\Delta m^2$ for 
$\nu_\mu \rightarrow \nu_\tau$ oscillations.
The 90\%  C.L. contour obtained by the Kamiokande
experiment is also shown.}
\label{fig:sk2}
\end{figure}

The zenith angle distributions for $e$-like and $\mu$-like 
FC and PC events are shown in Fig. \ref{fig:sk3}.
Note the near agreement of the $e$-like measured events with the MC
predictions without oscillations, and instead the deviations
from the no oscillation predictions for the $\mu$-like events.
Significantly small values of the double
ratio $R=(\mu/e)_{data}/(\mu/e)_{MC}$ in both the sub-GeV and
multi-GeV samples were obtained.
Several sources of systematic uncertainties in these measurement have
been considered, but none of these can alter significantly the
results.
Moreover, the $\mu$-like data exhibit a strong up/down 
asymmetry in zenith angle
($\Theta$) while no significant asymmetry is observed in the $e$-like data. 
The up/down ratio is expected to be near unity, really 
independent of the flux model
for $E_\nu > 1\ GeV$ (above this value the effects due to the Earth' magnetic
field on cosmic rays are small). The experimentally measured up/down
ratio is $0.54^{+0.06}_{-0.05} \pm 0.01$ for the multi-GeV FC+PC
$\mu$-like events.

SuperKamiokande estimated the oscillation parameters 
considering the $R$ measurements and the zenith
angle shapes separately. The 90\% CL
allowed regions for each case overlapped at $1 \times 10^{-3}
< \Delta m^2 < 4\times 10^{-3} \ eV^2$ for $sin^22\theta =1$.
Fig. \ref{fig:sk2}a shows the ratio of FC data to Monte Carlo for 
$e$-like and $\mu$-like events with $p>400\ MeV/c$
as a function of $L/E_{\nu}$, compared to the
expectation for $\nu_\mu \rightarrow \nu_\tau$ oscillations with
the best fit parameters from $R$ versus zenith angle. While the
$e$-like data show no significant variation in $L/E_{\nu}$, the
$\mu$-like events show a significant deficit at large $L/E_{\nu}$.
At large $L/E_{\nu}$, the $\nu_\mu$ have presumably undergone numerous 
oscillations
and have averaged out to roughly half the initial rate.

The final values for  $\nu_\mu \rightarrow \nu_\tau$ oscillations
are $sin^22\theta_{\mu\tau} =1$ and $\Delta m^2 = 0.0022\ eV^2$.
The contour plots for the neutrino oscillation parameters
for Kamiokande and SuperKamiokande are shown in Fig. \ref{fig:sk2}b.
Superkamiokande reported also data on up throughgoing muons, which 
agree with the
predictions with an oscillated flux with the above parameters\cite{koshiba}.

\section{Conclusions}

A wealth of new data on atmospheric neutrinos was presented
at the Neutrino'98 Conference.
The zenith angle distributions of atmospheric neutrino induced muons
differ in shape and in absolute value from the ones
predicted in the absence on neutrino oscillations.
In the vertical upgoing direction there are about 50\%
deficits for low and high energy muon events.
For $\nu_e$ induced  electrons there is no strong deviation from 
prediction. The ratio of muons to electrons normalized to the respective
MC predictions enhances the anomaly.

The new data are in agreement with the hypothesis of two flavour
$\nu_\mu \rightarrow \nu_\tau$ oscillations, with maximum mixing and
$\Delta  m^2\sim 0.0023\ eV^2$; the uncertainty in the 
$\Delta  m^2$ is relatively large.
The 90\% CL contours of MACRO and SuperKamiokande overlap  closely.

The present experiments 
on atmospheric neutrinos are disappearance experiments;
it would be nice to have a cross check with an  appearance experiment.

One cannot exclude $\nu_\mu \rightarrow \nu_s$ oscillations into a sterile
$\nu_s$. Answers to some detailed questions, like
the shape of the high energy muon angular
distribution may need more data.

\vskip .5cm
We would like to acknowledge the cooperation of many members of the 
MACRO collaboration, in particular of all the members of the neutrino working
group.

\end{document}